\documentclass[conference,10pt]{IEEEtran}

\usepackage{amssymb}
\usepackage{amsthm}

\usepackage[T1]{fontenc}
\usepackage{url}
\usepackage{ifthen}
\usepackage{cite}
\usepackage[cmex10]{amsmath} 

\usepackage{graphicx}

\usepackage{mathtools}

\usepackage[section]{placeins}

	\usepackage{cite}
	\usepackage[ruled,vlined,linesnumbered]{algorithm2e}

\usepackage{tikz}

\usepackage{verbatim}
\usetikzlibrary{arrows}

\newcommand{\Expect}{{\rm I\kern-.3em E}}

\begin{document}
\title{Deep Learning for Interference Identification:\\ Band, Training SNR, and Sample Selection}
\author{\large{Xiwen Zhang, {\em Student Member, IEEE}, Tolunay Seyfi, {\em Student Member, IEEE}},\\ \large{Shengtai Ju, {\em Student Member, IEEE}, Sharan Ramjee, {\em Student Member, IEEE}}, \\ \large{Aly El Gamal, {\em Member, IEEE} and Yonina C. Eldar, {\em Fellow, IEEE}}}

\maketitle

\begin{abstract}

\footnote{X. Zhang, T. Seyfi, S. Ju, S. Ramjee and A. El Gamal are with the Department of Electrical and Computer Engineering, Purdue University, West Lafayette, IN, 47907 USA (e-mail: zhan2977, tseyfi, ju10, sramjee, elgamala@purdue.edu).

Y. C. Eldar is with the Department of Math and Computer Science, Weizmann Institute of Science, Rechovot, Israel (e-mail: yonina@weizmann.ac.il).}We study the problem of interference source identification, through the lens of recognizing one of 15 different channels that belong to 3 different wireless technologies: Bluetooth, Zigbee, and WiFi. We employ deep learning algorithms trained on received samples taken from a 10 MHz band in the 2.4 GHz ISM Band. We obtain a classification accuracy of around 89.5\% using any of four different deep neural network architectures: CNN, ResNet, CLDNN, and LSTM, which demonstrate the generality of the effectiveness of deep learning at the considered task. Interestingly, our proposed CNN architecture requires approximately 60\% of the training time required by the state of the art while achieving slightly larger classification accuracy. We then focus on the CNN architecture and further optimize its training time while incurring minimal loss in classification accuracy using three different approaches: 1- Band Selection, where we only use samples belonging to the lower and uppermost 2 MHz bands, 2- SNR Selection, where we only use training samples belonging to a single SNR value, and 3- Sample Selection, where we try various sub-Nyquist sampling methods to select the subset of samples most relevant to the classification task. Our results confirm the feasibility of fast deep learning for wireless interference identification, by showing that the training time can be reduced by as much as 30x with minimal loss in accuracy.   
\end{abstract}
\section{Introduction}

With the omnipresence of wireless communication networks and their increasing demand for transferring wireless data, competition for scarce spectrum resources and interference management are becoming important problems. 
In particular, it is important to guarantee that different technologies, as well as different applications supported by the same technology, coexist and can simultaneously operate in an efficient manner. Examples of technologies that occupy overlapping frequency bands are WiFi, ZigBee, and Bluetooth, which share the 2.4 GHz ISM band. Identifying the technology and channel index of a received wireless signal is hence fundamental to the design of collaborative interference management schemes that allow for approaching the limits of wireless communications in next generation networks. In particular, intelligent use of the limited spectrum becomes an essential task for next-generation cognitive radio systems (see e.g., \cite{rajab-iwcmc}).


In \cite{mod-jsac18} and \cite{oshea-nn}, the feasibility of using deep learning to analyze received wireless communication signals was established, through the task of automatic recognition of the modulation type. It was shown, that proposed architectures such as Convolutional Long Short-Term Deep Neural Networks (CLDNN) \cite{sainath-icassp}, Long Short-Term Memory Neural Networks (LSTM)\cite{west-dyspan} and Deep Residual Networks (ResNet) \cite{oshea-ieee} lead to typical classification accuracy values around $90\%$ with 10 modulation types at high SNR.

The problem of Wireless Interference Identification (WII) is that of recognizing the technology of the received wireless signal, together with its channel index. In \cite{bitar-pimrc}, a 5-layer CNN architecture with a classification accuracy above 90\% at high SNR was proposed for identifying the technology of 802.x standards operating in the entire 80 MHz wide 2.4 GHz ISM band. 
In \cite{Schmidt-arXiv17}, a CNN-based architecture was proposed for the WII problem, based on sensing snapshots with a limited duration of $12.8~ \mu$s and a limited bandwidth of $10$ MHz. The dataset, that is analyzed by the CNN, consists of 15 classes representing packet transmissions of IEEE 802.11 b/g, IEEE 802.15.4 and IEEE 802.15.1 with overlapping frequency channels within the 2.4 GHz ISM band. It was shown that the CNN architecture surpasses state-of-the-art WII approaches with a classification accuracy greater than $95$\% for signal-to-noise ratios above -$5$ dB. 

In this work, we analyze the same dataset of \cite{Schmidt-arXiv17} with emphasis on reducing the training time than the state of the art, while maintaining high classification accuracy. We achieve an average accuracy of around $89.5$\% using four deep neural network architectures: CNN, ResNet, CLDNN, and LSTM. Also, the very high accuracy achieved in~\cite{Schmidt-arXiv17} at moderately high SNR values is maintained. We then proceed with optimizing the proposed CNN architecture. We chose the CNN architecture to create a fair comparison with the results in~\cite{Schmidt-arXiv17}, where only a CNN architecture is considered. It is interesting to note that our proposed CNN architecture requires 60\% of the training time required by the CNN architecture of \cite{Schmidt-arXiv17}, when we take the average result over 10 runs. We focus on three approaches to further reduce the training time of the proposed CNN architecture: 1- Band selection by analyzing the two lower and upper 2 MHz frequency bands, 2- Training SNR selection by selecting only a subset of the training set corresponding to one SNR value, and 3- PCA and various sub-Nyquist sampling techniques. The obtained results demonstrate the practical potential of fast deep learning for WII. 


\begin{table}
    \centering
    \caption{The considered 15 classes of channels.}
    \begin{tabular}{|c|c|c|c|}   
    \hline
    Class Index & Technology & Center Frequency & Channel Width \\    
    \hline
    1           & Bluetooth  & 2422 MHz         & 1 MHz  \\
    \hline
    2           & Bluetooth  & 2423 MHz         & 1 MHz  \\
    \hline
    3           & Bluetooth  & 2424 MHz         & 1 MHz  \\
    \hline
    4           & Bluetooth  & 2425 MHz         & 1 MHz  \\
    \hline
    5           & Bluetooth  & 2426 MHz         & 1 MHz  \\
    \hline
    6           & Bluetooth  & 2427 MHz         & 1 MHz  \\
    \hline
    7           & Bluetooth  & 2428 MHz         & 1 MHz  \\
    \hline
    8           & Bluetooth  & 2429 MHz         & 1 MHz  \\
    \hline
    9           & Bluetooth  & 2430 MHz         & 1 MHz  \\
    \hline
    10          & Bluetooth  & 2431 MHz         & 1 MHz  \\
    \hline
    11          & WiFi       & 2422 MHz         & 20 MHz \\
    \hline
    12          & WiFi       & 2427 MHz         & 20 MHz \\
    \hline
    13          & WiFi       & 2432 MHz         & 20 MHz \\
    \hline
    14          & Zigbee     & 2425 MHz         & 2 MHz  \\
    \hline
    15          & Zigbee     & 2430 MHz         & 2 MHz  \\
    \hline
    \end{tabular}
\end{table}

\section{Problem Setup}
We begin with a description of the considered problem and experimental setup.
We use the dataset generated by Schmidt et al. \cite{Schmidt-arXiv17}, which contains in total 225,225 sample vectors for 15 classes (as shown in Table I) in the SNR range of -20 dB to 20 dB with the step size of 2 dB\footnote{The dataset is available at \url{https://crawdad.org/owl/interference/20180925/}}. For the IEEE 802.11 b/g (WiFi) frames, the Physical Layer Mode is varied between CCK, PBCC, and OFDM.
For the IEEE 802.15.1 (Bluetooth) frames, the Transport Mode is varied between ACL, SCO, and eSCO.
For the IEEE 802.15.4 (Zigbee) frames, the ACK-frame is used.
For each class and each SNR value, there are 715 sample vectors, of which we use 480 for training and 235 for validation. Each sample vector consists of 128 I/Q samples, corresponding to $12.8 \mu s$. 
The I/Q samples of each sample vector are also transformed into the frequency domain by using the Fast Fourier Transform (FFT). Further, for the experiments of Section~\ref{sec:pca}, we obtained better accuracy and training time results by converting the I/Q samples into an Amplitude-Phase representation. For brevity, we refer to these datasets as FFT, I/Q and FFT Amp-Phase, respectively.


Given an input sample vector, the goal is to design deep neural network architectures that can be used to recognize the channel type, among the 15 classes shown in Table I. We tested four different architectures: CNN, ResNet, CLDNN, and LSTM. For all the four architectures, the Adam optimizer and a batch size of 256 are used in the training phase. For CNN, ResNet, and CLDNN, a learning rate of 0.0001 is used, while for the pure LSTM architecture, 0.001 is chosen for learning rate. Also, we use the Rectified Linear Unit (ReLU) as the activation function for all layers except the last one, for which we use the Softmax activation function. The Categorical Cross Entropy function is used for each as the loss function. For the CLDNN architecture, a dropout of 60\% is used for each layer except the last dense layer, while for the CNN architecture, a dropout of 60\% is used for each layer except the first convolutional layer and the last dense layer. For the first and second dense layers of ResNet, an Alpha Dropout of 10\% is used. We provide in Table \ref{table: neural network architecture} a summary of the four architectures. For the experimental setup, we used a GPU server equipped with an Nvidia Tesla P100 GPU and 16 GB of memory to train and test each deep neural network classifier. All reported training times are obtained by taking the average over 10 runs using the reported hardware\footnote{The code of this work is available at \url{https://github.com/dl4amc/dl4wii/}}.

\section{Results}

\begin{table*}
    \centering
    \caption{The considered 4 deep neural network architectures. The convolutional layer column indicates the number of feature maps and kernel size of each layer. The dense layer colunm indicates the input and output size of each layer.} 
    \begin{tabular}{|c|c|c|c|c|c|c|}   
    \hline
    Architecture & Activation Function & Convolutional Layer & Dense Layer & Recurrent Cells & Residual Stacks & Accuracy\\    
    \hline
    CNN \cite{Schmidt-arXiv17} & ReLU, Softmax & 64 $3*1$, 1024 $3*2$ & $126976*128$, $128*15$ &  &  & 0.8941 \\
    \hline
    CNN & ReLU, Softmax & 256 $3*1$, 256 $3*2$ & $31744*1024$, $1024*15$ &  &  & 0.8962          \\
    \hline
    LSTM   & ReLU, Softmax &  & $512*15$ & 512, 4 &  & 0.8965 \\
    \hline
    ResNet & ReLU, Softmax &  & $128*128$, $128*128$, $128*15$ &  & 5 & 0.8938  \\
    \hline
    CLDNN  & ReLU, Softmax & 256 $3*1$, 256 $3*2$ & $512, 256$, $256, 15$  & 256 &  & 0.8950 \\
    \hline
    \end{tabular}
    \label{table: neural network architecture}
\end{table*}

\begin{figure}
    \centering
	\includegraphics[width=160pt]{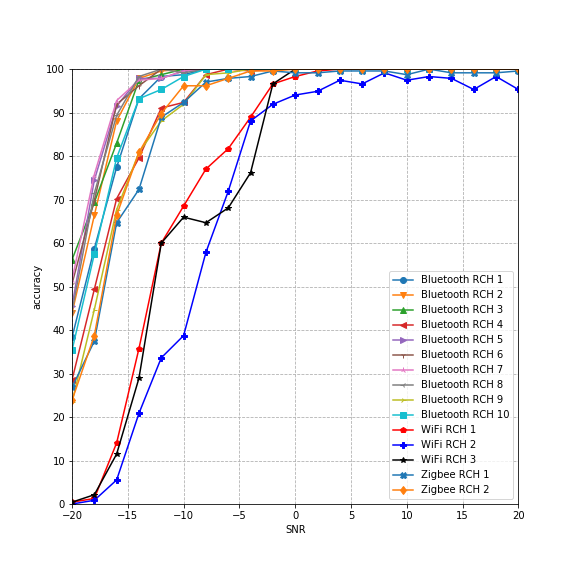}
	\caption{Classification accuracy vs. SNR of CNN on 10 MHz dataset.}
	\label{fig:cnn-line-graph}
\end{figure}

\begin{figure}
    \centering
	\includegraphics[width=200pt]{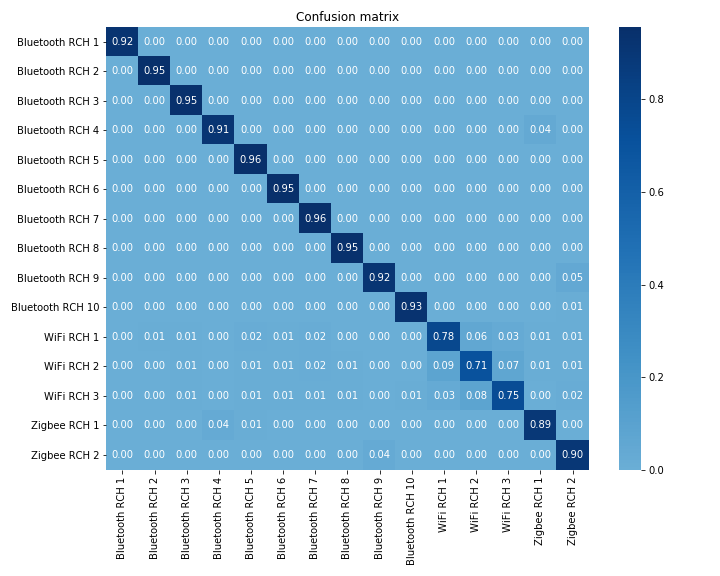}
	\caption{Average confusion matrix of CNN on 10 MHz dataset.}
	\label{fig:cnn-confusion-matrix}
\end{figure}

Our main objective is to reduce the required training time, while maintaining high classification accuracy. We first study four different architectures: CNN, ResNet, CLDNN, and LSTM. The design details of each architecture are listed in Table \ref{table: neural network architecture}. It is worth noting that all the reported results, except for those in Section~\ref{sec:pca}, are based on FFT I/Q data. We also notice that our proposed CNN architecture delivers a slightly higher accuracy than that of \cite{Schmidt-arXiv17}. Further, the average training time we obtained for our proposed CNN architecture is around 108s, as opposed to a 180s training time obtained for the original CNN architecture of \cite{Schmidt-arXiv17}.
In terms of classification accuracy, the results obtained for each wireless technology are very similar for all tested neural network architectures. 
Since the only architecture used in \cite{Schmidt-arXiv17} is a CNN, We focus on the CNN architecture for the remainder of this work, without expecting significant changes in accuracy if a different architecture is used.
As shown in Figure \ref{fig:cnn-line-graph}, for negative SNR dB values, the classification accuracy for the 3 classes of WiFi signals are significantly lower than those for Bluetooth or Zigbee signals.
Also, the confusion matrix in Figure \ref{fig:cnn-confusion-matrix} indicates that the reason for the poor performance on WiFi signals is not the difficulty in discriminating WiFi signals from Bluetooth or Zigbee, but the confusion between different channels of WiFi.
As shown in Table 1, there is a significant overlap between the bands occupied by different WiFi channels. Further, these bands are only captured in part by the sampled range of 2421.5-2431.5 MHz. We believe that these two reasons are causing the low classification accuracy between WiFi signals. We hence focus our band selection methods, presented in the following section, to capture bands close to WiFi center frequencies.



\subsection{Band Selection}\label{sec:band}
The objective of the experiments carried out here is to accelerate the training of deep neural network classifiers through band selection while retaining the classification accuracy as much as possible. We first introduce the definition of band selection and describe its influence on the dataset. Then, we compare the training time and accuracy before and after different choices of band selection. We summarize in Table \ref{table:band selection} the best set of results we have achieved through band selection, which demonstrates significant reductions of the training time while slightly impacting the classification accuracy.

By band selection, we mean utilizing only data from a subset of the original 10 MHz frequency range to train and test the neural network classifiers. After processing the time-domain dataset with FFT, band selection can be simply achieved through retaining parts of each frequency domain sample vector corresponding to the selected band. As the length of each sample vector gets shorter with band selection, the neural network architectures also shrink correspondingly. Note that band selection also results in fewer classes, as not all classes are observable in the narrower frequency range.

\begin{figure}
    \centering
	\includegraphics[width=160pt]{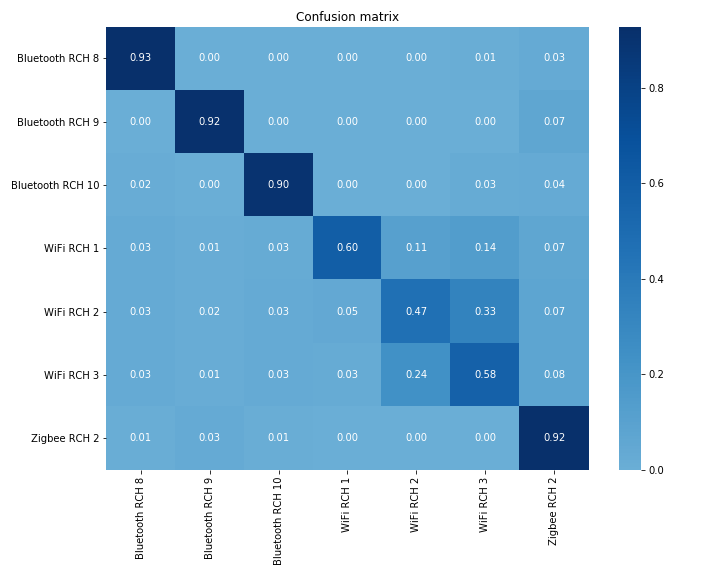}
	\caption{Average confusion matrix of CNN on 2 MHz dataset.}
	\label{fig:cnn-confusion-matrix-2429-2431}
\end{figure}

\begin{figure}
    \centering
	\includegraphics[width=160pt]{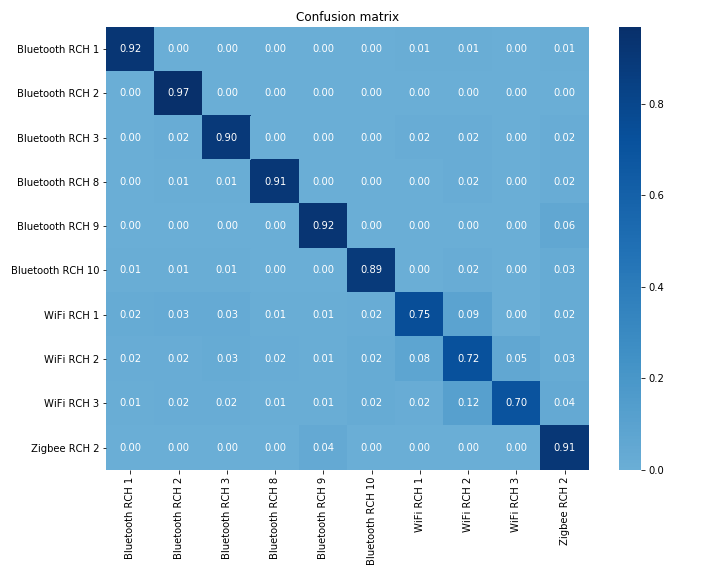}
	\caption{Average confusion matrix of CNN on 4 MHz dataset.}
	\label{fig:cnn-confusion-matrix-2422-2424-2429-2431}
\end{figure}

We start with selecting a narrow band of 2 MHz width. We choose the frequency range from 2429 to 2431 MHz as the selected sub-band. 
After this 2 MHz band selection, there are only 7 observable classes: classes 8, 9, and 10 (Bluetooth), classes 11, 12, and 13 (WiFi), and class 15 (Zigbee). The neural network architecture is the same as that before band selection, except that we add a dropout of 60\% after the first convolutional layer, which we found to have negligible influence on accuracy but makes the training process faster.
The resulting classification accuracy and training time are shown in Table \ref{table:band selection}. 
Compared with the classification accuracy achieved before band selection, we observe that the accuracy for WiFi signals is affected the most. 
Another interesting observation is from the confusion matrix, which is shown in Figure \ref{fig:cnn-confusion-matrix-2429-2431}. 
After band selection, the confusion between class 12 and class 13 is much more severe than that between class 11 and class 12 or between class 11 and class 13. 
However, in the confusion matrix before band selection, these numbers are comparable. 
We suspect that the reason is that the selected frequency range is just between, and close to, the center frequencies of classes 12 and 13, while the center frequency of class 11 is further from the selected band.

Based on the above insight, we select another 2 MHz band between 2422 and 2424 MHz to improve the WiFi classification accuracy. In the new 4 MHz wide frequency range (2422-2424 MHz and 2429-2431 MHz), 10 classes are observable, among which are 6 classes of Bluetooth, 3 classes of WiFi and 1 class of Zigbee.
The resulting classification accuracy and training time are shown in Table \ref{table:band selection}. We observe that after adding the other 2 MHz band, the accuracy of recognizing WiFi signals is significantly improved (from 0.5288 to 0.7253), which is near the performance before band selection. In fact, the accuracy obtained by the 4 MHz band selection is only slightly lower than the best results that have been achieved (The difference is smaller than 3\%). However, the training time is reduced by approximately 60\%.
We also tried another 4 MHz frequency range (2424-2426 MHz and 2429-2431 MHz). However, the accuracy of recognizing WiFi signal decreases from 0.7253 to 0.6363. We believe the reason is that the added band (2424-2426 MHz) is not close enough to the central frequency of any WiFi signal (2422, 2427, and 2432 MHz).

\begin{table}
    \centering
    \caption{Summary of classification accuracy and training time.}
    \begin{tabular}{|c|c|c|c|}   
    \hline
                              & 10 MHz & 4 MHz  & 2 MHz   \\    
    \hline
      Bluetooth               & 0.9402 & 0.9196 & 0.9149  \\
    \hline
      WiFi                    & 0.7467 & 0.7323 & 0.5255  \\
    \hline
      Zigbee                  & 0.8918 & 0.8967 & 0.9286  \\
    \hline
      Total Training Time     & 108.04s & 65.096s & 40.745s \\
    \hline
      Number of Epochs        & 6.6    & 15.8   & 28.1   \\
    \hline
      Training Time per Epoch & 16.37s  & 4.12s   & 1.45s   \\
    \hline
    \end{tabular}
    \label{table:band selection}
\end{table}
\FloatBarrier

\subsection{Training SNR Selection}\label{sec:snr}
\begin{figure}
    \centering
	\includegraphics[width=\linewidth]{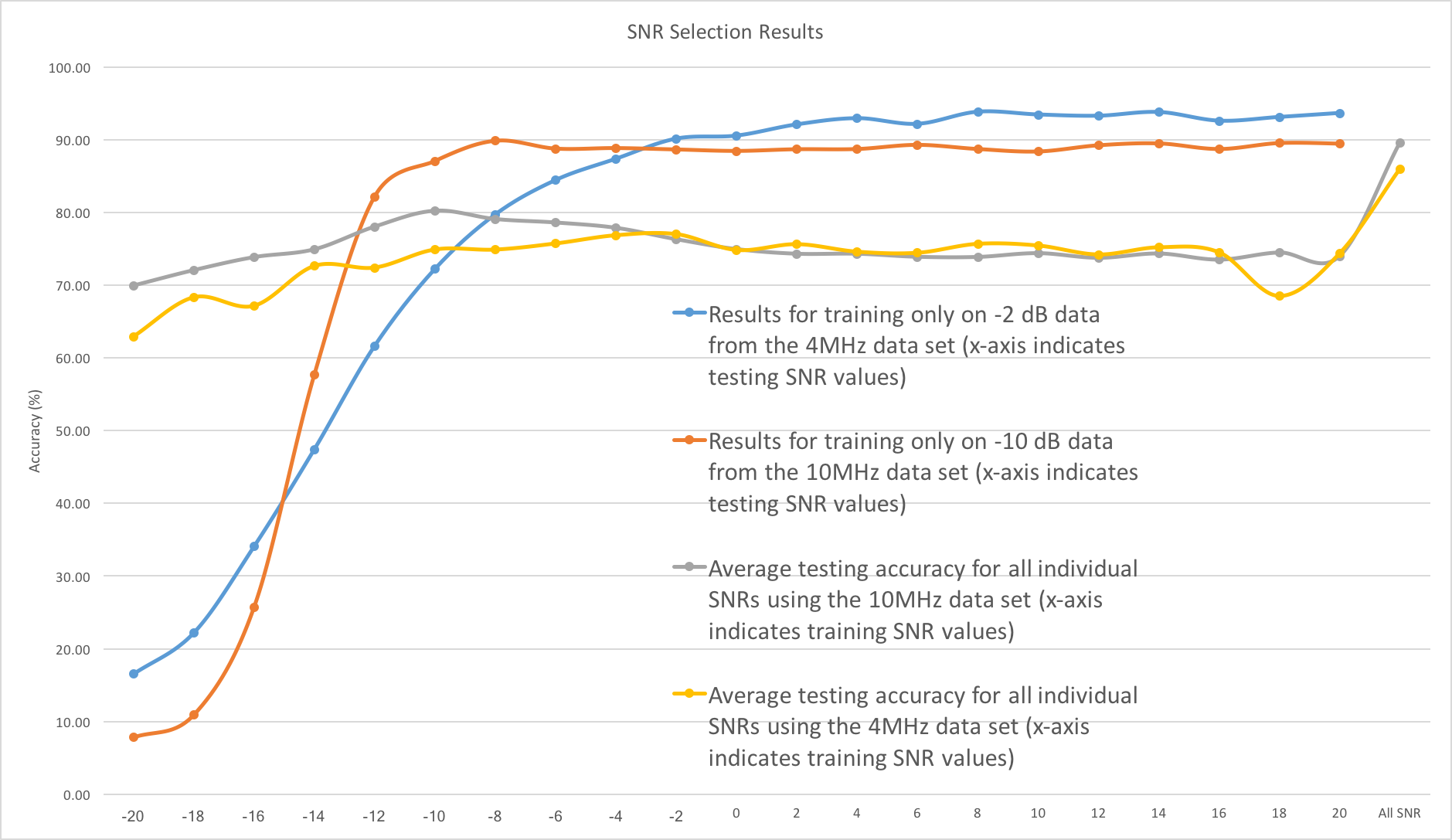}
	\caption{SNR Selection Results}
	\label{fig:snr_sel}
\end{figure}

In this section, we discuss training the CNN architecture mentioned in Section II with FFT I/Q datasets collected at a single SNR value for both the full 10 MHz dataset and the 4 MHz dataset introduced in Section~\ref{sec:band}. The training time is reduced drastically while maintaining high classification accuracy for testing data with high SNR values. The testing accuracy versus training SNR values for the 10MHz dataset is shown in Figure \ref{fig:snr_sel}. As shown in the figure, the average testing accuracy for different training SNR values seems close and has an average accuracy of approximately 75\% compared to an average testing accuracy of 90\% using the entire training dataset. Training with -10 dB data gave us the best average classification accuracy among all training SNR values, with an overall accuracy of slightly over 80\%. When training with -10 dB data, the classification accuracy for high SNR testing data decreased, while accuracy for low SNR testing data increased, compared to training with high SNR data. The major confusions are across the WiFi signals. The testing accuracy for each SNR value while training with -10 dB is shown in Figure \ref{fig:snr_sel}. The training time per epoch for using the entire 10 MHz training dataset is 16.37 seconds. The training time per epoch is reduced to 0.984 second using training data at only one SNR value. The total number of epochs is shown in Table \ref{tab:snr_label}. By training our model with only a single SNR value, the total training time is reduced by approximately 92.3\%. Figure \ref{fig:snr_sel} also shows the average testing accuracy for all individual training SNR values using the 4 MHz dataset. The overall average testing accuracy across all training SNR values is around 73\% compared to an accuracy of 86\% when trained with data that contains all SNR values. For the 4 MHz dataset, training with only -2 dB data led to the best performance across all available training SNR values, with an accuracy of approximately 77\%. Figure \ref{fig:snr_sel} shows the testing accuracy for each SNR value with the model trained only on -2 dB data. The classification accuracy for high SNR values is well above 90\%. The total training time was reduced by approximately 90.9\% as shown in Table \ref{tab:snr_label}. Overall, SNR selection could lead to reducing the total training time by more than 18x, while maintaining the classification accuracy for high SNR testing data above 90\%. 
\begin{table}
    \centering
    \caption{Training Time and Number of Epochs for SNR Selection.}
    \begin{tabular}{|c|c|c|c|}
    \hline
                & Time per Epoch & Number of Epochs &Accuracy\\
    \hline
     All SNR 10 MHz     &16.37s  & 6.6  &0.8962\\
    \hline
     -10 dB 10 MHz      &0.984s   &8.5  &0.8022\\
    \hline
     All SNR 4 MHz      &4.12s   &15.8  &0.8614\\
    \hline
     -2 dB 4 MHz        &0.61s   &9.7  &0.77\\
    \hline
    \end{tabular}
    \label{tab:snr_label}
\end{table}

\subsection{PCA and Sample Selection}\label{sec:pca}
\begin{figure}
	\includegraphics[width=1\linewidth]{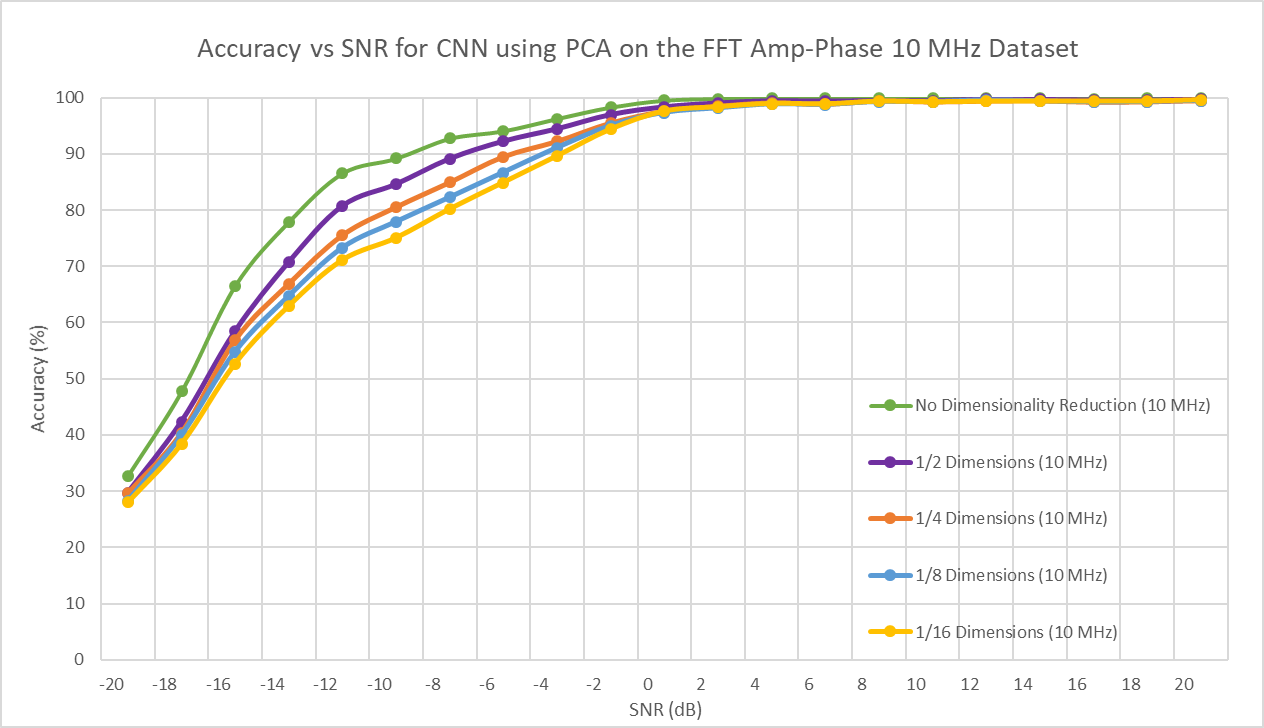}
	\caption{PCA on FFT Amp-Phase data}
	\label{fig:pca_fft_amp_phase}
\end{figure}

\begin{figure}
	\includegraphics[width=1\linewidth]{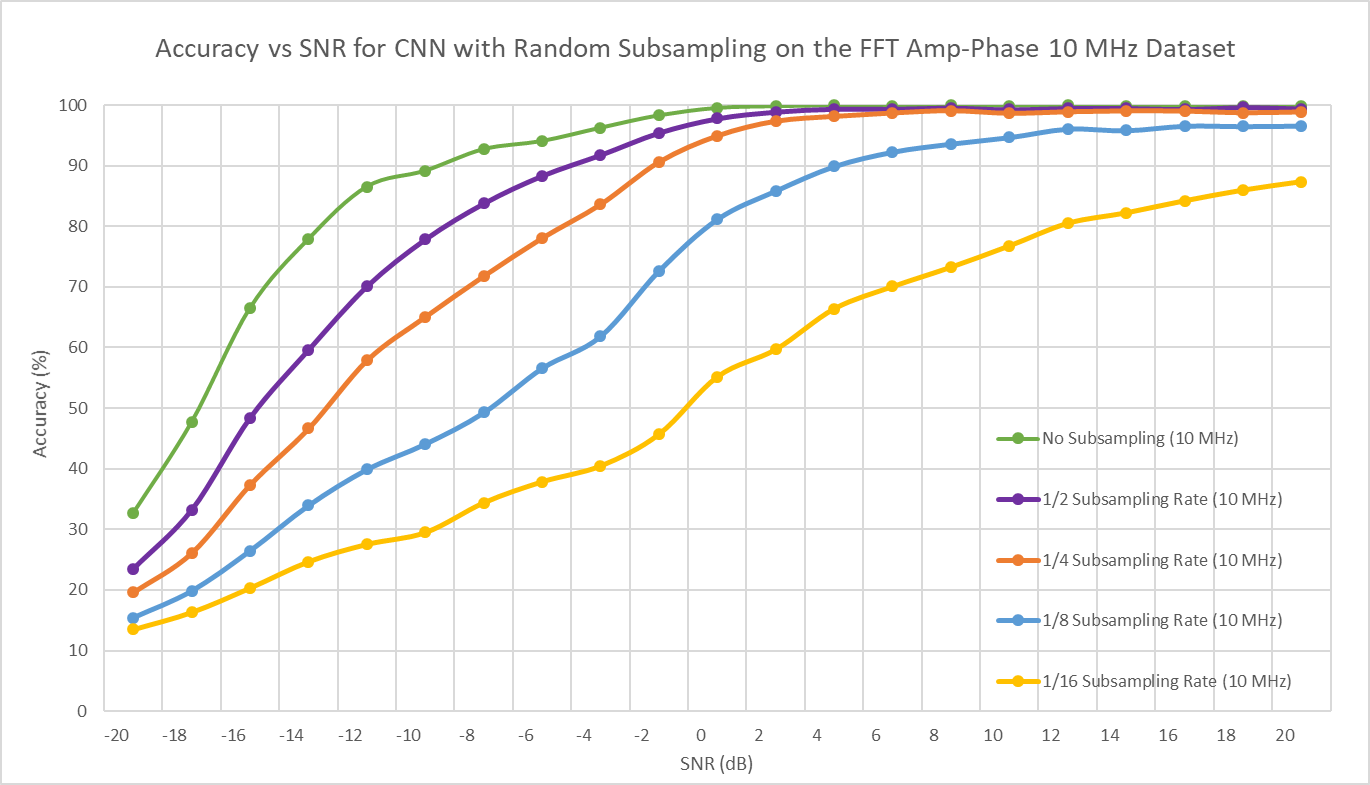}
	\caption{Random Subsampling on FFT Amp-Phase data}
	\label{fig:random_fft_amp_phase}
\end{figure}

Finally, we discuss the effects of using Principal Component Analysis (PCA) and various subsampling techniques on the training time and classification accuracy of the considered CNN. Although we experimented with PCA, Random subsampling, Uniform subsampling, and High Magnitude Rank subsampling (see \cite{mod-jsac18} for descriptions) for each of the 10 MHz and 4 MHz datasets, we only highlight here interesting results.


\begin{table}
    \caption{Comparison of Training Times using PCA}
    \begin{tabular}{|c|c|c|c|}
    \hline
            Dimensions/Samples & Time per Epoch & Epochs & Accuracy \\
    \hline
     All (10 MHz)      &16.37s   &6.6    &0.8962 \\
    \hline
     1/2 (10 MHz)      &7.79s    &8.6    &0.8726 \\
    \hline
     1/4 (10 MHz)      &3.86s    &8.5    &0.8576 \\
    \hline
     1/8 (10 MHz)      &2.16s    &7.4    &0.8487 \\
    \hline
     1/16 (10 MHz)     &1.78s    &7.3    &0.8411 \\
    \hline
     All (4 MHz)       &4.12s    &15.8   &0.8614 \\
    \hline
     1/2 (4 MHz)       &2.72s    &12.1   &0.8358\\
    \hline
     1/4 (4 MHz)       &1.64s    &8.6    &0.8310\\
    \hline
     1/8 (4 MHz)       &1.33s    &8.2    &0.8220\\
    \hline
    \end{tabular}
    \label{tab:pca_sample_selection}
\end{table}

\begin{figure}
	\includegraphics[width=1\linewidth]{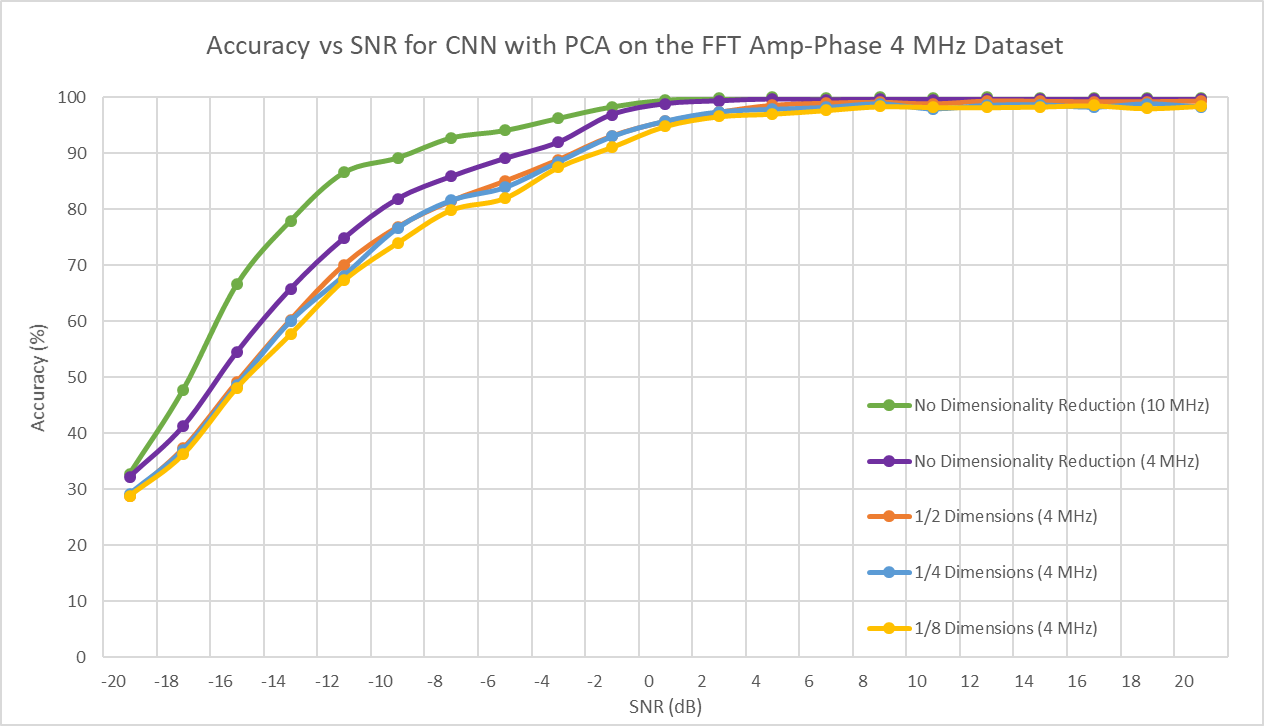}
	\caption{PCA on FFT Amp-Phase 4 MHz data}
	\label{fig:pca_fft_amp_phase_subset}
\end{figure}


In Figure \ref{fig:pca_fft_amp_phase}, we demonstrate the effect of PCA for the FFT Amp-Phase data. It is interesting to observe that negligible loss in accuracy is incurred for SNR values above 0 dB and a compression rate as high as 16x. In Figure \ref{fig:random_fft_amp_phase}, we demonstrate the results obtained through Random subsampling of the FFT Amp-Phase data, and observe large drops in accuracy at low SNR values compared to PCA. However, we still observe impressive results at high SNR values with negligible loss in accuracy for a subsampling rate as low as $\frac{1}{4}$. It is worth mentioning that we also obtained similar results with Uniform subsampling. Finally, we obtained interesting results with a combination of band selection and dimensionality reduction. In Figure~\ref{fig:pca_fft_amp_phase_subset}, we depict the results for PCA. It is interesting to notice the robustness of the classification accuracy at moderately high SNR values, with a combination of band selection and a reduction rate of up to 8x, which reduces the training time significantly as shown in Table~\ref{tab:pca_sample_selection}. We also obtained a similar insight regarding the robustness of the classification accuracy with band selection and Random subsampling. In general, there is a proportional drop in the training time as can be observed in Table \ref{tab:pca_sample_selection} when the number of dimensions/samples is reduced using PCA. We note that the training times for Random and Uniform subsampling are similar to those listed in Table \ref{tab:pca_sample_selection}.

\section{Concluding Remarks}
We first presented in this work four deep neural architectures that lead to an average accuracy around 89.5\% when classifying 15 different channels occupying the 2.4 GHz ISM band, using a narrow band of 10 MHz. We further demonstrated how to significantly reduce the training time with minimal loss in classification accuracy, through band and training SNR selection, as well as PCA and sample selection through various sub-Nyquist sampling methods. One particular result indicates that selecting a band of 4 MHz that captures the center frequencies of key wide channels, as well as selecting a single optimized SNR value for training (-2 dB), leads to reducing the total training time by 30x (compared to \cite{Schmidt-arXiv17}), while incurring negligible loss in classification accuracy for testing SNR values above -4 dB.    
	
\bibliographystyle{IEEEtran}

\end{document}